\documentclass[aps,prd,twocolumn,tightenlines]{revtex4}
\usepackage{epsfig}
\usepackage{graphicx}
\usepackage{psfrag}
\usepackage{amsmath,amssymb,psfrag,slashed,graphicx}
\usepackage{colordvi}
\usepackage{color}
\usepackage{url}
\usepackage{slashed}
\usepackage{color}
\usepackage[normalem]{ulem}
\usepackage[colorlinks,urlcolor=blue]{hyperref}

\newcommand{\bra}[1]{\left<#1\left|}
\newcommand{\ket}[1]{\right|#1\right>}
\newcommand{\braket}[1]{\left<#1\right>}

\begin{document}

\title{Proton mass decomposition: naturalness and interpretations}

\author{Xiangdong Ji}
\email{xji@umd.edu}
\affiliation{Center for Nuclear Femtography, SURA, \\ 1201 New York Ave. NW, Washington, DC 20005, USA}
\affiliation{Department of Physics, University of Maryland, College Park, MD 20742, USA}

\date{\today}

\begin{abstract}
I discuss the scope and naturalness of the proton mass decomposition (or sum rule) published
in PRL74, 1071 (1995)
and answer a few criticisms that appeared recently in the literature, focusing particularly on its interpretation and the quantum anomalous energy contribution. I comment on the so-called frame-independent or invariant-mass decomposition
from the trace of the energy-momentum tensor. I stress
the importance of measuring the quantum anomalous energy through
experiments. Finally, I point out a large discrepancy in the scalar
radius of the nucleon extracted from vector-meson productions and lattice QCD calculations.
\end{abstract}

\maketitle

\section{introduction}
In 1995, I wrote a paper that discussed for the first time a
proton mass decomposition (or sum rule) in quantum chromodynamics (QCD)~\cite{Ji:1994av}.
The goal was to express the proton mass, or that of any other hadron, in terms of
quark and gluon energies corresponding to the matrix elements measurable
in experiments, in Einstein's spirit that inertial mass is a store of
energy~\cite{Hecht:2011}. It received relatively little attention
in the community for more than 20 years. Recently, understanding the proton mass has become an important topic in hadronic physics, particularly in light of new experiments at Jefferson Lab 12 GeV facility and future Electron-Ion Collider~\cite{Dudek:2012vr,Accardi:2012qut}.
Three workshops have been organized by a group of mass enthusiasts
on this topic~\cite{Meziani}. In the process of understanding better
about the proton mass structure and its experimental verification,
a number of questions concerning the naturalness and interpretations of my decomposition have emerged.  The apparent discrepancies claimed in the literature have been reduced to the question whether the anomalous contribution to the nucleon mass is an unambiguously defined physical quantity which can be
calculated on the lattice and ultimately be determined experimentally. Following and expanding the arguments given in~\cite{Ji:1994av}, I will stress that the answer is affirmative.
In doing so I will discuss in detail alternative proposals and will explain why I do not think that they are helpful to better understand the relevant physics.

In section II, I review the original derivation, emphasizing the key point that mass is the rest energy and there exists a complete energy basis to express the mass in QCD. In section III, I discuss why there is a quantum anomalous energy contribution and comment on its natural appearance in QCD Hamiltonian through time dilatation.
In Section IV, I consider a well-known relation involving the matrix element of the trace of the QCD energy-momentum tensor, arguing
it is not a natural frame-independent mass decomposition, but rather
about scale symmetry breaking effects.
In section V, I discuss the so-called ``pressure contribution'' to the mass sum rule and argue that it is based on a questionable picture. Consideration of such an effect contradicts the well-known concept of
the quark mass contribution to the proton mass. In Sec. VI, I make some comments
about the heavy quark contribution and the structure of the pion mass. In Sec. VII,
I discuss the mass distribution recently
emphasized by Kharzeev~\cite{Kharzeev:2021qkd}, pointing out the
difference between scalar and mass radii. Finally, I conclude in Sec. VIII.

\section{Original derivation: symmetry and complete energy basis}

While mass is a frame-independent concept, to calculate the mass
for a composite system like the proton, it is best to take the working definition as the energy in the rest frame~\cite{Okun:1989}.
If one chooses a general frame, the energy will
contains both the mass effect and kinetic energy, and one has to find ways
to subtract the latter to obtain the pure mass. Such a subtraction is
neither illuminating nor fun, and hardly produce any additional insight about the nature of mass. See Sec. IV for more discussion.

Therefore, the starting point of Ref. \cite{Ji:1994av} is
simply to write the rest-frame energy of the proton
in terms of quark and gluon energies. One can of course work out
the QCD Hamiltonian from the action and examine all the terms one gets, and
ponder what they mean in terms of quarks and gluons. However,
Ref. \cite{Ji:1994av} classifies the operators in terms of Lorentz symmetry, a tensor method that has been universally used in field
theories (see for example Ref.~\cite{Buchalla:1995vs} to study the
Hamiltonian for weak decay where flavor symmetry is the key),
similar to the Wigner-Eckart theorem in quantum mechanics.
Symmetry maximally protects physics interpretations of individual parts because ultra-violet (UV) divergences introduce the scheme and scale dependence and make the physical meaning of a decomposition not as clear-cut as in classical physics. In fact, the result of Ref. \cite{Ji:1994av} shows that
the QCD Hamiltonian contains four independent terms with {\it three combinations scheme independent}, and hence genuinely physical: They are the quark mass, anomalous energy and total tensor contributions.

Here is a simple review of the derivation: The QCD energy-momentum tensor (EMT) is the conserved currents associated with the space-time translational symmetry from Noether's theorem. It is
a rank-2 tensor $T^{\alpha\beta}(x)$ and the conservation law requires
\begin{equation}
\partial_\alpha T^{\alpha\beta}(x)=0 \ .
\end{equation}
Furthermore, it can be improved by the Belinfante-Rosenfeld process, and hence
the EMT can also be assumed to be symmetric. A symmetric rank-2 tensor can in
general be {\it uniquely} decomposed as
\begin{equation}
\begin{split}
    T^{\alpha\beta}(x)=& \bar T^{\alpha\beta}(x) +\hat T^{\alpha\beta}(x) \ ,\\
\end{split}
\end{equation}
with
\begin{equation}
    \hat T^{\alpha\beta}(x)\equiv \frac{1}{4}g^{\alpha\beta} T^{\rho}_{\rho}(x)\ .
\end{equation}
The trace-less part $\bar T^{\alpha\beta}(x)$ and the trace part $\hat T^{\alpha\beta}(x)$ correspond to irreducible spin-2 and spin-0 representations of Lorentz group, respectively. They have different Lorentz transformation and renormalization properties. The traceless part of the EMT in massless QCD (we ignore the quark masses here for simplicity, but will return to them later)
can be written as
\begin{equation}
    \bar T^{\alpha\beta}(x)=\bar T_q^{\alpha\beta}(x,\mu)+ \bar T_g^{\alpha\beta}(x,\mu)\ ,
\end{equation}
where we can express either in terms of the un-renormalized or renormalized traceless quark and gluon tensors with $\mu$ indicating
renormalization scheme and scale dependence (remembering the well-known
twist-2 operators entering deep-inelastic scattering cross section~\cite{Peskin:1995ev}),
\begin{align}
    \bar  T_q^{\alpha\beta}(\mu)&\equiv \left(\bar \psi \gamma^{(\alpha} i\overleftrightarrow{D}^{\beta)} \psi\right)_{\rm{R}}(\mu) \ ,\\
    \bar  T_g^{\alpha\beta}(\mu)&\equiv \left(-F^{\alpha \rho}F^{\beta}_{\;\;\rho}+\frac{1}{4}g^{\alpha\beta}F^2\right)_{\rm{R}}(\mu) \ ,
\end{align}
with $()$ symmetrizing all the indices, $\overrightarrow{D}^\mu=\overrightarrow{\partial}^\mu-igA^\mu$, $\overleftarrow{D}^\mu=\overleftarrow{\partial}^\mu+igA^\mu$ and $\overleftrightarrow{D}^\mu=1/2(\overrightarrow{D}^\mu-\overleftarrow{D}^\mu)$. The $x$ dependence is suppressed hereafter. Then the total QCD EMT can be decomposed as
\begin{equation}
\label{Tdecomp}
     T^{\mu\nu}= \bar T_q^{\mu\nu}(\mu)+\bar T_g^{\mu\nu}(\mu) +\hat T^{\mu\nu} \ .
\end{equation}
where the trace term is scheme and scale independent as the whole EMT is.

The matrix elements of above operators are constrained by their Lorentz transformation properties, for instance for the total QCD EMT
\begin{equation}
    \bra{P} T^{\mu\nu} \ket{P}=2P^\mu P^\nu \ ,
\label{emtme}
\end{equation}
where the covariant normalization $\langle P|P\rangle = 2E\delta^3(0)$ is used, differing
from the original paper by $1/2M$ with $M$ as the proton mass.
For the quark and gluon trace-less EMT
\begin{equation}
\label{Tqgdecomp}
    \bra{P} \bar T^{\mu\nu}_{q,g} \ket{P}=2\braket{x_{q,g}} \left( P^\mu P^\nu -\frac{1}{4}g^{\mu\nu} M^2\right)\ ,
\end{equation}
with $\braket{x_{q,g}}$ can be measured as the momentum fractions carried by quarks and gluons in
the infinite momentum frame and satisfy the momentum sum rule,
\begin{equation}
    \braket{x_{q}}(\mu)+\braket{x_g}(\mu)=1\ .
\end{equation}
On the other hand, the trace part of the EMT satisfies
\begin{equation}
\label{Tadecomp}
    \bra{P} \hat T^{\mu\nu} \ket{P}=\frac{1}{2}g^{\mu\nu} M^2\ .
\end{equation}
which is nothing but the well-known trace relation~\cite{Shifman:1978zn},
\begin{equation}
    \langle P|T^\alpha_\alpha|P\rangle = 2M^2 \ ,
\label{emtrace}
\end{equation}
directly from Eq. (\ref{emtme}).

To relate the EMT decomposition to the mass sum rule, consider the Hamiltonian $ H$ that gives the energy of states, which can be expressed as the charge of the conserved EMT as
\begin{equation}
\begin{split}
     H_{\rm QCD} &\equiv \int \text{d}^3\vec x ~T^{00}(\vec x,t=0)\ ,
\end{split}
\end{equation}
where $t=0$ will be suppressed. Then the mass of the nucleon is simply the energy in the rest frame
\begin{equation}
    M\equiv \bra{\vec P=0}  H\ket{\vec P=0}/\braket{\vec P=0|\vec P=0} \ .
\end{equation}
The expression on the right hand side will be denoted as $\braket{H}$. Using eq. (\ref{Tdecomp}), it then becomes
\begin{equation}
    M=\int \text{d}^3 \vec x \braket{\bar T_q^{00}(\mu)+\bar T_g^{00}(\mu) +\hat T^{00}}\ .
\end{equation}
Defines the decomposition of Hamiltonian as:
\begin{equation}
\begin{split}
    H_q(\mu) &\equiv  \int \text{d}^3 \vec x~ \bar T^{00}_{q}(\vec {x},\mu)\ , \\
    &=  i\int \text{d}^3 \vec x \left(\bar \psi \vec \gamma\cdot \overleftrightarrow{\vec D} \psi\right)_{\rm{R}}(\vec {x},\mu)\ ,
\end{split}
\label{quarkenergy}
\end{equation}
\begin{equation}
\begin{split}
    H_g(\mu) &\equiv  \int \text{d}^3 \vec x ~\bar T^{00}_{g}(\vec {x},\mu) \ ,\\
    &=  \frac{1}{2}\int \text{d}^3 \vec x \left(\vec E^2+\vec B^2\right)_{\rm{R}}(\vec {x},\mu) \ ,
\end{split}
\label{emenergy}
\end{equation}
and~\cite{Collins:1976yq, Nielsen:1977sy}
\begin{equation}
\begin{split}
    H_a &\equiv  \int \text{d}^3 \vec x ~\hat T^{00}(\vec {x})\ , \\
    &=  \frac{\beta(g)}{8 g(\mu)}\int \text{d}^3 \vec x \left (F^{\alpha\beta}F_{\alpha\beta}\right)_{\rm{R}}(\vec {x})\ ,
\end{split}
\end{equation}
which is scale-independent and generates the quantum anomalous energy (QAE).
The total Hamiltonian is
\begin{equation}
\begin{split}
   H_{\rm QCD}=H_q(\mu)+H_g(\mu)+H_a\ .
\end{split}
\end{equation}
The scale dependence in $H_q$ and $H_g$ cancels.

With the matrix elements in eq. (\ref{Tqgdecomp}) and eq. (\ref{Tadecomp}), the mass can be written with
\begin{equation}
\begin{split}
    M_q(\mu) \equiv \braket{H_q(\mu)}= \frac{3}{4}M \braket{x_q}(\mu) \ ,
\end{split}
\end{equation}
\begin{equation}
\begin{split}
    M_g(\mu) \equiv \braket{H_g(\mu)}= \frac{3}{4} M \braket{x_g}(\mu) \ ,
\end{split}
\end{equation}
and
\begin{equation}
\begin{split}
    M_a \equiv \braket{H_a}= \frac{1}{4} M \ .
\end{split}
\end{equation}
Then the total mass can be written as a sum of three terms
\begin{eqnarray}
    M&=&  M_q+M_g+M_a  \nonumber
    \\ &=& \frac{3}{4} M \braket{x_q}(\mu) +\frac{3}{4} M \braket{x_g}(\mu) +\frac{1}{4} M \ .
\end{eqnarray}
If the quark masses are non-zero, one ends up with a decomposition with four terms, each
of which are related to experimental observables and calculable in lattice QCD~\cite{Chen:2005mg,Abdel-Rehim:2016won,Alexandrou:2017oeh,Yang:2018nqn,He:2021bof}.
The largest error comes from the anomalous energy calculation~\cite{He:2021bof}, and
the state-of-art verification of the above sum rule is at 10\% level.
A discussion of the relevant physics has been made recently in instanton vacuum picture in which the anomalous energy is related to the compressiblity of the instantons~\cite{Zahed:2021fxk}. While the $ M_q(\mu)$ and $ M_g(\mu)$ can be measured in deep inelastic scattering,
$ M_a$ is related to the matrix element of $F^2$ in the proton which can be potentially
probed through a color dipole in experiments such as threshold $J/\psi$ photo or
electro production~\cite{Kharzeev:1998bz,Hatta:2018ina,Mamo:2019mka,Wang:2019mza,Boussarie:2020vmu, Meziani:2020oks}. The form factor of the anomalous energy can also be measured through deeply virtual Compton scattering.

\section{Quantum anomalous energy contribution}

It must be a bit surprising to notice for the first time that
there is a new source of energy coming from the quantum trace anomaly
in field theories~\cite{Collins:1976yq, Nielsen:1977sy}.
After all, the energy sources in QCD are the same as those in
quantum electrodynamics (QED): there is an energy from
electromagnetic fields similar to what's in Eq. (\ref{emenergy})
from the traceless EMT.
The electron's kinetic energy is also well-known through solving relativistic Dirac equation in an external Coulomb
field in advanced quantum mechanics. The electron's mass contribution
to energy is associated with the scalar density $\bar\psi\psi$
which becomes the number of electrons in non-relativistic limit.
All these exist in classical electromagnetism. On the other hand, the
quantum anomalous energy (QAE) is a new source which exists only
in quantum theory of fields, and hardly has been discussed in the literature in the past. It arises from the breaking
of scale symmetry due to UV divergences in quantum field theories.

One could argue the derivation of the previous section is
mathematically correct, and so QAE must be there.
Or, the QCD Hamiltonian
along with its various energy sources is completely determined
once the action is known, and one just need to go through the renormalization
and show that it is present. However, there is a
physically more enlightening way to reveal its existence~\cite{Ji:2021pys}.

Once there is an action in space and time, the conserved energy is
associated with its time-translation symmetry. One can actually {\it derive}
the Hamiltonian as the response of the action under
a time-dependent translation, $t\rightarrow t+\delta \lambda t$, which amounts
a re-scaling in time~\cite{DiFrancesco:1997nk},
\begin{equation}
S\rightarrow S+\delta\lambda \int dt H \ .
\end{equation}
For example, for the pure gauge theory, in terms of the time re-scaled field variables $A_{\mu}'(t',\vec{x})=A_{\mu}(t,\vec{x}) (1+\delta^{\mu 0}\delta \lambda)$, the action transforms to
\begin{align} \label{eq:assy}
S'=\frac{1}{2g_0^2}\int d^4x' \bigg(-\frac{1}{1+\delta \lambda} ({\vec B}')^2+(1+\delta \lambda)(\vec{E}')^2\bigg) \ ,
\end{align}
where $g_0$ is the bare coupling.
The contribution at order $\delta \lambda$ is nothing but the well-known canonical
Hamiltonian in Eq. (17).

However, in quantum theory, one needs a regulator to regularize
the UV divergences, resulting in the scale dependence or running of
the couplings. In a class of regulators which break the time-scaling
symmetry such as the lattice cutoff, the running of couplings can be
induced through such a breaking. Then a new and anomalous
contribution to the Hamiltonian appears naturally, although the anomaly itself is
regulator independent.
With such a regulator, $g_0$ in Eq.~(\ref{eq:assy}) depends on $\delta \lambda$~\cite{Karsch:1982ve},
the new term in the Hamiltonian corresponds to the derivative of $g_0$
with respect to $\delta \lambda$~\cite{Rothe:1995av},
\begin{align}\label{eq:anoaH}
H_a=\frac{1}{4}\int d^3 \vec{x}\frac{\beta(g_0)}{2g_0}F^2 \ ,
\end{align}
which reflects the logarithmic running of couplings $g_0$ with the UV
cut-off scale through the beta function $\beta(g_0)$. If quarks have masses, there is also a term with the mass anomalous dimension $\gamma_m$. The above agrees with Ref.~\cite{Ji:1994av} where this term
is derived from the trace anomaly.
In regulators which do not break time-translation symmetry such as
dimensional regularization, the anomaly term is hidden in the classical
form of the Hamiltonian.

Therefore, QED does have an anomalous term in the Hamiltonian~\cite{Ji:1998bf}.
Since the covariant field theory formulation does not involve identifying
the Hamiltonian, the role of the anomalous energy is implicit. However, the
QAE contribution to the Lamb shift in the hydrogen atom has been calculated recently~\cite{Sun:2020ksc,Ji:2021pys}.

The Hamiltonian in Ref.~\cite{Metz:2020vxd} does not contain the explicit
anomalous term due to the use of dimensional regularization~(see Eq. (53-56)).
However, it is not absent but hidden in Eq. (55). Their definition of the renormalized $\int d^3x (E^2+B^2)_R$ is from the full gluon EMT which contains the trace anomaly in quantum theory.
If one uses the trace-less EMT to define the gluon energy as in the classical case~\cite{Luke:1992tm,Ji:1994av,Kharzeev:1995ij},
the anomaly term in their Hamiltonian becomes obvious. As a further confirmation of
the anomaly in Ref.~\cite{Metz:2020vxd}, the matrix element of Eq. (55) is in Eq. (76) which contains the $1-b$ from the anomaly ($x=0$ and $y=\gamma_m$), after implementing Eq. (69). The sum of three contributions in Ref.~\cite{Metz:2020vxd} exactly reproduces the full result in Ref.\cite{Ji:1994av}.

As explained in Ref.\cite{Ji:2021pys,Ji:1994av}, the anomaly contribution is scale
independent in chiral limit, and contains the key information about the scale
generation of the proton mass, reflected through a Higgs-like mechanism. As is manifest in Eq. (69),
it has the same role in providing an independent scale to the proton mass
as the quark masses, which have been recognized and explicitly
separated in Eq. (54) of Ref.~\cite{Metz:2020vxd}. In the instanton liquid picture of the QCD vacuum~\cite{Shuryak:1999fe},
the QAE is related to the QCD vacuum compressibility~\cite{Zahed:2021fxk}; in the MIT bag model, it is related to the scalar energy density or the bag constant inside the nucleon~\cite{Chodos:1974je}.
Combining the novel QAE contribution to the proton mass together with the
ordinary $\int d^3x (E^2+B^2)_R$ contribution, which is measurable by a different experiment and is scheme dependent due to mixing with quarks, likely misses the fundamental insight on the origin of the proton mass.
See further comments in Sec. V.

\section{Toward Lorentz-invariant mass decomposition}

Since mass is a scalar, a frame-independent understanding
might be more fundamental~\cite{Roberts:2016mhh}. However, this appears
misleading. For a massive particle, even for a pseudo Goldstone boson like pion,
the rest frame is special, where all energies are internal.
When the particle moves, it has kinetic
energy, $K=E-m/c^2$ (take $c=1$). The relation $m^2= (K+m)^2 - P^2$
ensures the mechanical energy is properly subtracted so that only
the internal energy as mass is left. In this view, a frame-independent concept of mass
only implies a proper subtraction procedure for the kinetic contribution,
and does not necessarily provide a better insight.
Nonetheless, it is worth to investigate if a frame-independent
subtraction can be made and, at the same time, any additional
feature can be learned along the way.

A natural starting point is the QCD mass Casimir $\hat P^\mu\hat P_\mu$ (from now on I use
an hat on a symbol to denote an operator, and that without is an eigenvalue). Take
its expectation value in the nucleon state as the {\it definition} of frame-independent mass,
\begin{equation}
             M^2 = \frac{\langle P|\hat P^\mu \hat P_\mu |P\rangle}{\langle P|P\rangle} \ .
\label{covamass}
\end{equation}
Note that this expression is independent of the state normalization.  Some insight
on the invariant mass may be gleaned from studying the general operator structure of the Casimir.
One can proceed by expressing one or both
$\hat P_\mu$ as quark and gluon operators: if one replaces both
operators by kinetic eigenvalues, one ends up with a totality.
On the other hand, if both $\hat P^\mu$ remain
as dynamical operators, one ends up with a very complicated expression which may not be very
illuminating. No one has gone down this path in the literature.

A more interesting approach is to replace one of $\hat P^\mu$ as kinetic
eigenvalue and expand the other as dynamical operator. This is attractive because
in the rest frame, it naturally reduces to the matrix element of the QCD Hamiltonian as in Sec. II. In the general frame, one has the matrix element
of a linear combination of the Hamiltonian and momentum operators,
\begin{equation}
    M = \langle \gamma (H - \vec{\beta}\cdot \vec{P})\rangle \ ,
\label{generalH}
\end{equation}
where $\gamma, \vec{\beta}$ are Lorentz-boost factors.
The linear combination above is just the Lorentz transformation
of the Hamiltonian operator in the nucleon's rest frame. One can
use the tensor decomposition as in Sec. II to write
it in terms of three different contributions (again ignore the quark mass),
\begin{equation}
    M = \langle \gamma (H_q-\vec{\beta}\cdot \vec{P_q})\rangle  +
    \langle \gamma (H_g-\vec{\beta}\cdot \vec{P_g})\rangle  + \gamma\langle H_a\rangle
\end{equation}
The combination $\gamma H_q-\vec{\beta}\cdot \vec{P_q}$ gives
a contribution independent of frame,
\begin{eqnarray}
 \langle \gamma (H_q-\vec{\beta}\cdot \vec{P_q})\rangle
 &=& \langle x\rangle_q M
 \left[\gamma(\gamma - 1/(4\gamma))-\gamma^2\beta^2\right] \nonumber \\
 &=& \frac{3}{4}\langle x\rangle_q M
\end{eqnarray}
It shall be remarked that the classical electromagnetic field
contribution to the electron mass obeys exactly this 3/4 rule under Lorentz transformation~\cite{Feynman:1963uxa}.
On the other hand, $\gamma\langle H_a\rangle $ is a scalar and contribute $M/4$.
Therefore the decomposition in Sec. II has {\it a Lorentz-invariant
interpretation}. In the infinite momentum frame, Eq. (\ref{generalH})
reduces to the light-front Hamiltonian $P^-$ plus a kinetically-suppressed
$P^+$ term. The decomposition of light-front energy
$P^-$ in terms of twist-four operators have been studied in~\cite{Ji:2020baz,Hatta:2020iin}.
The fact that the infinite momentum frame mass contains both kinematic
and dynamic contributions is similar to the proton's transverse spin
which contains both twist-2 and twist-3 sum rules~\cite{Guo:2021aik}.

In the literature, it has often been stated that
there is a frame-independent mass sum rule
\begin{equation}
\label{covamasstrace}
 2M^2 = \left\langle P\left|(1+\gamma_m)m\bar\psi\psi + \frac{\beta(g)}{2g}F^2\right|P\right\rangle \ ,
\end{equation}
which is Eq. (\ref{emtrace}) except with the explicit trace operator~\cite{Shifman:1978zn}.
The operator is a scalar and here the state is normalized
covariantly $\langle P|P\rangle = 2E (2\pi)^3\delta^3(0)$.
However, as I argue, {\it one cannot find a natural connection} between the above equation and the invariant mass operator in Eq. (\ref{covamass}).
I will try to go through the derivation below, but
there a number of arbitrary steps in the middle,
and hence one cannot claim that it provides a natural insight about the nucleon mass, other than that it provides one of many qualities involving the mass.

To derive Eq. (\ref{covamasstrace}) from the mass operator,  one first splits
up the energy from the momentum,
\begin{equation}
  M^2 = \langle \hat H^2 \rangle - \langle \hat{P}^2_i\rangle \ ,
\end{equation}
where $\langle... \rangle $ is a short hand with normalization factor in the
denominator taken into account. One can replace one of $\hat H$ by its eigenenvalue
$E$, and keep the other as operator expressed in term of $T^{00}$
\begin{eqnarray}
       \langle H^2 \rangle &=& E\left\langle \int d^3 x T^{00}\right\rangle \nonumber \\
                         &=& EV\langle T^{00}\rangle \nonumber  \\
                         &=& \langle P|T^{00}|P\rangle/2\ ,
       \label{energysquare}
\end{eqnarray}
where $V = (2\pi)^3 \delta^3(0)$ and in the final line the state is normalized covariantly.

%Here one may ask
%why shall one treat two Hamiltonian factor differently? One can also write
%$\langle H^2 \rangle = E(\alpha\langle H\rangle + (1-\alpha)E)$, or $\langle H^2 \rangle
%= E^{1-\alpha} \langle H^{1-\alpha}\rangle $ with $\alpha$ as arbitrary parameter to derive
%more equality.

Now in order to get a trace as in Eq. (\ref{covamasstrace}), one cannot substitute the
QCD operator for $\hat P_i$ directly in there, because it
is directly related to $\hat T^{0i}$ not $\hat T^{ii}$. Therefore,
one has to first replace $\hat P^i$ by its eigenvalue, and then
replace the eigenvalue by another matrix element through
the $(ii)$ component of Eq. (\ref{emtme}),
\begin{equation}
                  2(P_i)^2 = \langle P|T^{ii}|P\rangle \  .
\label{momentumsquare}
\end{equation}
These substitutions are mathematically correct but do not make a lot of sense
in physics. It certainly is not an explanation what the momentum is.
Combining Eqs. (\ref{energysquare}) and (\ref{momentumsquare}),
one gets Eq. (\ref{covamasstrace}). Since the above derivation involves
arbitrary steps, the result is hardly an explanation
what the invariant mass is.

My view is that the above equation simply expresses a relationship.
One can derive other similar relations, such as
\begin{equation}
    M =\frac{4}{3} \left(M_q(\mu)+M_g(\mu)\right) \ .
\end{equation}
These relations do not provide a direct explanation on the nature of mass as
the sum of all parts. In fact there is something unusual
about Eq. (\ref{covamasstrace}): an explanation of mass depends on the normalization
of a state. If one uses a different normalization, the relation
changes and then one needs to explanation why there is an extra
factor. One of course can argue that the covariance fixed
the factor of $E$, but still one could have an arbitrary numerical factor
and then one has to carefully fixed the factor through energy conservation
in the original equation.

Eq.~(\ref{covamasstrace}), however, is useful in two senses.
First of all, it expresses a relation between mass-squared
and the matrix element of scale symmetry breaking terms.
In the chiral limit, there is only one scalar, $F^2$, that breaks the scale symmetry and
sets the scale of QCD. Then the equation expresses
the fact that QCD relates all dimensional quantities to this scale.
The nucleon mass is non-vanishing because of the presence of this symmetry breaking term.
Like the spectrum of hydrogen atoms, as well as chemical and biological energies, all
are proportional to the mass of the electron. If the electron mass were to vanish,
all these energies vanish. But the electron mass does not give an explanation
why the hydrogen spectrum is the way it is. It is the proportional
coefficient that is controlled by the underlying physics.
When the quarks are massive, there is an additional
scale breaking term $m\bar\psi\psi$.
The relative magnitudes of the quark mass and anomaly terms
are fixed through their importance as the scale-symmetry breaking sources
in a hadron.

Quantitatively, Eq. (\ref{covamasstrace}) express the proton mass entirely in terms of
the scalar energy. The nucleon mass contains both scalar and tensor energies, $M=E_T+E_S$~\cite{Ji:2021pys}.
Using a virial theorem $E_T=3E_S$, one can replace the tensor energy in favor of the scalar one,
resulting
\begin{equation}
   M = 4E_S \ ,
\end{equation}
where $E_S$ is the scalar energy including the quark masses and is frame independent.
As in a harmonic oscillator where $E=T+V$, $\langle T\rangle = \langle V\rangle$,
the meaning of the above equation or Eq. (\ref{covamasstrace}), is the
similar to equations $\langle E\rangle = 2 \langle T\rangle$
or $\langle E\rangle = 2 \langle V\rangle$ for the oscillator.

Therefore, Eq. (\ref{covamasstrace}) is neither a frame-independent explanation
of the proton mass, nor a mass sum rule because the individual terms cannot be
identified properly as the proper components of the mass. Nonetheless, it is an interesting and useful relation.

Finally, it has been suggested that the anomaly in $T^\mu_{\mu}$
can be split into contributions from the quark and gluon parts
of the EMT separately~\cite{Hatta:2018sqd,Tanaka:2018nae}.
This splitting is scheme dependent: even in the context
of dimensional regularization, it depends on if the separate trace
obtained from the unrenormalized EMT or the renormalized version although
the sum is the same. Physical meaning
of the separation, including the martix elements and contributions to the mass, is unclear as both parts
depend on the same set of renormalized operators.

\section{Pressure Effect in Proton Mass?}

As explained in Sec. II, the goal of Ref.~\cite{Ji:1994av} was modest:
identifying distinct and complete energy operators in the QCD Hamiltonian, and relate their contributions to the proton mass
to experimental observables. I did not try to come up with a model or a picture for these independent contributions other than to take at face value of the operator structures. Because the nucleon wave function is complicated and unknown, any attempt for
a serious interpretation of the expectation values must deem incomplete. However, the decomposition is solid in QCD, as there are no superfluous operators present. Moreover, there are many ways to arrive at the same result, although the tensor decomposition is the quickest.

In Ref. \cite{Lorce:2017xzd}, the author appears to misunderstand the intent and content of Ref.~\cite{Ji:1994av},
%see for example the statement after Eq. (49), ``He... implicitly adopted an effective coupled four-fluid picture of the hadron.'' The author
claiming that the ``the pressure effects'' have been overlooked.
%. Based
%on this mis-understanding,
The author advocated ``a semi-classical picture'', ``an effective two-fluid picture...described in terms of their own energy`
densities and pressures'', arriving at a surprising conclusion
that ``the so-called quark mass and trace anomaly contributions (to mass) appear
to be {\it purely conventional}.''

There is a large gap between the expectation value of the Hamiltonian in the ground state of a quantum many-body system and a fluid model. In the
former case,  it is a result involving complicated quantum wave
function which entangles all degrees of freedom. In the latter, the energy-momentum tensor
is used as dynamical variables of the system after certain local averages
can be meaningfully made. For example, in the classical textbooks of Landau~\cite{landau},
conditions have been explicitly stated for an effective description in terms
of pressure and internal energy. One of those is that the particle's
mean free path must be much smaller than volume elements which
are taken as fundamental degrees of freedom. In QCD, only at high-temperature
and density, a fluid description of the combined quark and gluon plasma
might make sense~\cite{Shuryak:2014zxa}.

%Realizing the nucleon is not ``a set of perfect fluids'', the author of
%Ref. \cite{Lorce:2017xzd} nevertheless pressed on with ``the semiclassical limit'' which
%was not defined in the paper, and used ``the terminology of continuum mechanics,
%such as internal energy and pressure'', which is absent in
%field theory textbooks.

Physics models aside, what Ref. \cite{Lorce:2017xzd} actually criticised
is the tensor decomposition shown in Sec.II. The author
compared the expectation value of $\langle P|T^{\mu\nu}|P\rangle$
with the EMT of a perfect fluid, and identified the contribution
of $T^{ii}$ as ``mechanical pressure''.  Therefore the author concluded that
tensor decomposition mixes in the pressure
contribution in internal energy. Considering the $T^{00}$ operator, and
if one makes a traceless and traceful separation, one writes
\begin{equation}
     T^{00} = \frac{1}{4}\left(3T^{00}+T^{ii} \right) + \frac{1}{4}\left(T^{00}-T^{ii} \right)
     \ .
\end{equation}
In this equation, it appears that one artificially introduces
the stress tensor operator $T^{ij}$, which would smirch the physics
interpretation of two separate terms, as $T^{ii}$ does not seem to
have to do with the ``internal energy'' of the system.

While such criticism might be true in general, it cannot be lodged against
an analysis at a fundamental level.
First of all, the tensor decomposition does not introduce any new operator
structures. It is simply a method to group the terms already present
in the Hamiltonian. Endowing the quantum operators
in a field theory with the meanings of physical observables
in a thermal system generates exotic interpretations:
For instance, the energy momentum tensor of the
quarks is $T^{\mu\nu} = \bar \psi \gamma^{\mu}iD^{\nu}\psi$.
Following Ref. \cite{Lorce:2017xzd}, Dirac equation
\begin{equation}
  i(\gamma^0 D^0-\gamma^iD^i-m)\psi = 0\ ,
\end{equation}
shall be interpreted as ``internal-energy/pressure'' balance equation,
with the mass term as the residue. Moreover,
in the relativistic Hamiltonian
\begin{equation}
  H = \psi^\dagger (i\vec{\alpha}\cdot \vec{D} + m\beta)\psi \ ,
\end{equation}
the kinetic energy comes from the pure ``pressure''!
Unfortunately, these interpretations hardly bring in any new
insight: ``pressure'' and ``internal energy'' add no additional
physical significance at the fundamental level.

Adopting the logic of Ref. \cite{Lorce:2017xzd}, one can only talk about
the separate quark and gluon ``internal energy'' contributions to mass without scalar and
tensor separation, $M=U_q+U_g$.  According to Eq. (57-60), the author suggested
that only the following combinations have no ``pressure effect'',
\begin{eqnarray}
    U_q &=&  \langle H_q\rangle = M_q + M_m = A_q(0) + \bar C_q(0) \nonumber \\ &=& \left[\frac{3}{4}A_q(0) + \left(\frac{1}{4}A_q(0)+ \bar C_q(0)\right)\right]M  \ , \\
    U_g &=&  \langle H_g\rangle = M_g + M_a = A_g(0) + \bar C_g(0) \nonumber \\ &=& \left[\frac{3}{4}A_g(0) + \left(\frac{1}{4}A_g(0)+\bar C_g(0)\right)\right]M \ ,
\end{eqnarray}
where $A_{q,g}(t)$ and $\bar C_{q,g}(t)$ are form factors defined in Ref. \cite{Ji:1996ek}.
In the second lines of the above equations, we have shown the trace-less and trace separation.
It is unclear what new insight can be brought to
the understanding of the proton mass through the process of regrouping if any.
To the contrary, this rearrangement stands to lose much:
\begin{itemize}
\item{The decomposition of the mass in $U_q(\mu)$ and $U_g(\mu)$ is less physical due to
complicated scheme dependence. In fact, according to Ref. \cite{Hatta:2018sqd}, $\bar C_q(\mu)$
and $\bar C_g(\mu)$ can have additional scheme dependence, and they mixes among themselves, rendering interpretation of $U_q$ and $U_g$ less straightforward.}
\item{One can no longer isolate the quark mass contribution to the proton mass,
which is scale invariant and has been studied for many decades in chiral perturbation theory~\cite{Gasser:1984gg}. It can also be measured through experiments and calculable through lattice QCD~\cite{Yang:2018nqn,Alexandrou:2019brg}}.
\item{One can no longer isolate the anomaly contribution to the proton mass which is a result of scale symmetry breaking due to UV fluctuations. It is scale invariant and can be measured experimentally and calculable independently on the lattice~\cite{He:2021bof}. The quark mass and quantum
anomalous energy have the same status according to Eq.(\ref{covamasstrace}).}
\end{itemize}
Therefore, while the four term decomposition in Sec. II tells us a lot about the proton mass structure, the suggestion made in Ref. \cite{Lorce:2017xzd} lead to the unfortunate conclusion that the scale and scheme independent quark mass and quantum anomaly contributions are ``purely conventional''. This shall miss some of the most important physics about the proton mass.

\section{Heavy Quark and Pion}

The heavy quark contribution to the proton mass has been
computed on lattice, particularly for the charm quark~\cite{Gong:2013vja,Alexandrou:2019brg}.
The nominal contribution is large: $\langle P|m_c \bar cc|P\rangle$ is on the order of 100 MeV.
However, the anomaly will cancel it mostly~\cite{Shifman:1978zn}. One estimation
of the anomaly contribution associated with charm flavor
is on the order of $-65$ MeV. Therefore the total contribution from
charm quark is around 35 MeV, perhaps smaller than the strange quark contribution. The bottom quark contribution
is smaller, and the top will be entirely negligible.

Finally, I would like to make a comment on the mass structure of the pion.
Its mass decomposition using the same method has been considered in Ref. \cite{Ji:1995sv}, where it was found that
the QAE contributes about 1/8 of the pion mass
\begin{equation}
   \langle H_a\rangle_\pi = m_\pi/8 \ .
\end{equation}
Therefore the gluon part of the wave function shall approach that of the QCD
vacuum in the chiral limit. Likewise, the standard
gluon energy contribution also vanishes
in this limit,
\begin{equation}
   \langle H_g\rangle_\pi \sim 3m_\pi/8 \ .
\end{equation}
for the experimental data on the parton momentum distribution.
While the quark mass contributes roughly 1/2 of the mass
through chiral perturbation theory, the quark
kinetic and potential energy contribution is small
\begin{equation}
   \langle H_q\rangle_\pi \sim 0 \ .
\end{equation}
Therefore, in the chiral limit all contributions
vanish smoothly without any fine tuning.
The above results provide important constraints on any realistic picture of the
pion mass.

\section{Mass, scalar and tensor Radii}

Three nucleon gravitational form factors
can be defined
from the full QCD energy-momentum tensor~\cite{Pagels:1966zza,Ji:1996ek},
\begin{equation}
\begin{split}
\label{Tmunuformfactors}
& \bra{P'}T^{\mu \nu}\ket{P}=\bar{u}\left(P^{\prime}\right)\bigg{[}A\left(Q^{2}\right) \gamma^{(\mu} \bar{P}^{\nu)}\\  &+B\left(Q^{2}\right) \bar{P}^{(\mu} i \sigma^{\nu) \alpha} q_{\alpha} / 2 M \\ &
\;\;+C\left(Q^{2}\right)\left(q^{\mu} q^{\nu}-g^{\mu \nu} q^{2}\right) / M
 \bigg{]} u(P)\ ,
\end{split}
\end{equation}
where $q^\mu=P'^\mu-P^\mu$ is the four-momentum transfer, $\bar P^\mu = (P'^\mu+P^\mu)/2$,
and $Q^2=-q^2$. All three $A(Q^2)$, $B(Q^2)$, and $C(Q^2)$ are scheme and scale independent, as the electromagnetic Dirac and Pauli form factors, and hence are completely physical.
The above equation indicates the traceless and trace parts of form factors
are closely related by energy-momentum conservation.
The trace part defines
the dilatation or scalar form factor,
\begin{equation}
\bra{P'}T^\mu_ \mu\ket{P}=\bar{u}\left(P^{\prime}\right)
u(P)G_s(Q^2) \ ,
\end{equation}
where,
\begin{equation}
    G_s(Q^2)=\left[ MA(Q^2) -B(Q^2)\frac{Q^2}{4M}
+C(Q^2)\frac{3Q^2}{M}\right] \  .
\end{equation}
In the chiral limit, the scalar form factor
can be obtained from the trace anomaly contribution~\cite{Kharzeev:1995ij}.
On the other hand, the (00) component defines
the {\it mass form factor} in the Breit frame,
\begin{equation}
\bra{P'}T^{00}\ket{P}=\bar{u}\left(P^{\prime}\right)
u(P) G_m(Q^2) \ .
\end{equation}
where
\begin{equation}
    G_m(Q^2)=\left[ MA\left(Q^{2}\right) -B(Q^2)\frac{Q^2}{4M}
+C(Q^2)\frac{Q^2}{M}\right] \ .
\end{equation}
All form factors $A$, $B$ and $C$ can be in principle measured from
the deep-exclusive processes like deeply-virtual Compton scattering through sum rules derived for generalized parton distributions $E$ and $H$~\cite{Ji:1996ek,Ji:1997gm}.

For large $M$, the scalar and form factors
are the same and are related to $A(Q^2)$ only, as stressed in~\cite{Kharzeev:2021qkd}. However, for finite $M$, the two form factors
are different. If defining the radii
in terms of
\begin{equation}
    \langle r^2 \rangle_{s,m}
     = -6 \frac{dG_{s,m}(Q^2)/M}{dQ^2} \ ,
\end{equation}
we have
\begin{eqnarray}
        \langle r^2 \rangle_{s}
     &=& -6\frac{dA(Q^2)}{dQ^2}
     - 18 \frac{C(0)}{M^2} \ , \nonumber \\
        \langle r^2 \rangle_{m}
     &=& -6\frac{dA(Q^2)}{dQ^2}
     - 6 \frac{C(0)}{M^2}\ ,
\end{eqnarray}
where we used the fact that $B(0)=0$ ~\cite{Pagels:1966zza,Ji:1996ek}.
So the difference between scalar and
mass radius is
\begin{equation}
    \langle r^2 \rangle_{s} -
    \langle r^2 \rangle_{m} = -12 \frac{C(0)}{M^2} \ ,
\label{difference}
\end{equation}
where $C(0)= C_q+C_g$ is a sum of the quark and gluon contributions.

Recently, there have been extractions
of the scalar radius from $J/\psi$, $\phi$ and $\omega$ productions
near threshold using Vector Dominance Model (VDM)~\cite{Kharzeev:2021qkd,Wang:2021dis}. Neglecting the finite quark
mass effect in the trace of QCD EMT, it was found from the GlueX $J/\psi$ data that~\cite{Kharzeev:2021qkd}
\begin{equation}
     \langle r^2 \rangle_{s} = (0.55 \pm 0.03~\rm fm)^2 \ ,
    \label{expr1}
\end{equation}
and from $\phi$ and $\omega$ data~\cite{Wang:2021dis},
\begin{equation}
     \langle r^2 \rangle_{s} = (0.64 \pm 0.03~\rm fm)^2 \ .
         \label{expr2}
\end{equation}
Both results are smaller than the charge radius. Alternatively, it has been suggested that the GlueX data is related to $A$ form factor only in the holographic model ~\cite{Mamo:2021krl}, which produces a radius $\langle r^2\rangle_A = (~0.61\rm fm)^2$.

There have been lattice QCD calculations
of the quark form factors $A_q(0)$ and $C_q(0)$~\cite{Hagler:2007xi,Hagler:2009ni} and more recently the gluon ones~\cite{Shanahan:2018pib} from the traceless part of the EMT. Although
the quark masses in these works are somewhat heavier compared
with the physical ones (with pion mass  450$\sim$ 490 MeV), it is interesting
to extract the scalar and mass radii
from these calculations. Since $A_g$ is softer
than $A_q$, the radius from $A$-form factor
is dominated by the gluon contribution,
\begin{equation}
      \langle r^2\rangle_A =  -6\frac{dA(Q^2)}{dQ^2} = (0.47~\rm fm)^2
\end{equation}
where the quark contribution includes only the connected contractions from up and down quarks.
Meanwhile, $C(0)$ has a considerable uncertainty but a large central value,
\begin{equation}
      C(0) \equiv (D_q(\mu) + D_g(\mu))/4 \sim -1.25 \ ,
\end{equation}
where the scheme and scale dependence cancels in the sum. This yields,
\begin{equation}
     \langle r^2 \rangle_{s} \sim (1.1 \rm ~fm)^2 \ ,
\end{equation}
which is considerably bigger than those in Eqs.~(\ref{expr1},\ref{expr2}), and might be reasonable if considered as a bag radius
in the MIT bag model~\cite{Chodos:1974je}.
The mass radius is
\begin{equation}
 \langle r^2 \rangle_{m} \sim (0.74 \rm ~fm)^2,
\end{equation}
a bit smaller than the charge radius. If these lattice results are taken seriously,
they might imply that the VDM extractions
have a large unknown systematics.

On the other hand, if the sign of $C(0)$ is definitely negative~\cite{Polyakov:2018zvc}, the scalar radius must be larger than the mass radius.
This is reasonable because it is generally expected that the $0^{++}$ scalar excitation has a lower mass than $2^{++}$ which also contributes to the mass form factor~\cite{Chen:2005mg,Mamo:2021krl}. Assuming the mass radius is not negative, Eq. (\ref{difference}) puts a constraint
\begin{equation}
      \langle r^2 \rangle_{s} \ge -12 \frac{C(0)}{M^2} \ .
\end{equation}
If one uses $C(0)= -1.25$, the right hand side
is $(0.81~ \rm fm)^2$. If the extracted experimental value is trust-able, it clearly violates the above bound. Then the gluonic $C(0)$ from lattice should be much smaller than that in Ref.~\cite{Shanahan:2018pib}. Indeed, if one uses $(0.55~ \rm fm)^2$ as the upper limit for the $C(0)$, one has
\begin{equation}
    |C(0)|\le 0.57
\end{equation}
which constrains $|D_q(0) + D_g(0)|\le 2.3$, a result can be checked in future high-precision lattice calculations.

Finally, one can also define a tensor radius
$ \langle r^2\rangle_t$ through (see Eq. (37))
\begin{equation}
     \langle r^2\rangle_m =\frac{3}{4} \langle r^2\rangle_t + \frac{1}{4}
     \langle r^2\rangle_s  \ .
\end{equation}
We have then,
\begin{equation}
       \langle r^2 \rangle_{t}
     = -6\frac{dA(Q^2)}{dQ^2}
     - 2 \frac{C(0)}{M^2}\ ,
\end{equation}
The lattice data give
\begin{equation}
    \langle r^2 \rangle_t =(0.58~\rm fm)^2 \ .
\end{equation}
One can also see
\begin{equation}
     \langle r^2 \rangle_{s}
      -  \langle r^2 \rangle_{t}
       = -16 \frac{C(0)}{M^2} \ ,
\end{equation}
and thus $C(0)$ also determines
the difference between scalar and tensor
radii. Assuming the dipole form of the form factors, $G_{t,s}(t) = M/(1-t/m_{t,s}^2)^2$, one gets the following relation between the scalar and tensor masses,
\begin{equation}
       \frac{1}{m_s^2} - \frac{1}{m_t^2} = -\frac{4C(0)}{3M^2}
\end{equation}
The lattice data yield: $m_s = 620$~MeV, and $
m_t = 1.18$~MeV.

\section{conclusion}
In this paper, I argue that the most natural way to study the mass structure of the proton is through its rest frame energy. It contains four terms with three combinations being scheme independent and thus physical, and the decomposition can be made frame independent or Lorentz covariant.
Other variants of this decomposition, two terms or three terms, are discussed but appear less natural. Moreover, the so-called frame-independent mass structure from the EMT trace equation does not appear to provide additional insight on the mass.
Measuring the anomaly contribution experimentally is important, which may help to explain the origin of the proton mass through a Higgs-like mechanism. Finally, I comment on the discrepancy in the scalar
radius of the nucleon extracted from vector-meson production and lattice QCD calculations.

{\it Acknowledgment.}---We thank K. F. Liu, Y. Z. Liu, D. Kharzeev, A. Metz, Z.-E. Meziani, B. Pasquini, S. Rodini, A. Schaefer, F. Yuan, and I. Zahed for discussions and comments on the manuscript, and Y. Guo and Y. Su for help. I particularly thank Y. Z. Liu for helping with the gravitational form factors. This material is supported by the U.S. Department of Energy, Office of Science, Office of Nuclear Physics, under contract number DE-SC0020682, and by Southeastern Universities Research Association.

\bibliographystyle{apsrev4-1}
\bibliography{bibliography}

\end{document}